\documentclass[aps,prl,twocolumn,superscriptaddress]{revtex4}
\usepackage{amsmath}
\usepackage[colorlinks]{hyperref}
\usepackage{amssymb}
\usepackage{color}
\usepackage{graphicx,amsmath}
\UseRawInputEncoding \input
\hypersetup{colorlinks,citecolor=red,linkcolor=blue,urlcolor=blue}

\begin{document}

\title{Comment on: "The future of the correlated electron problem", arXiv: 2010.00584}
\author{V. R. Shaginyan}\email{vrshag@thd.pnpi.spb.ru} \affiliation{Petersburg
Nuclear Physics Institute of NRC "Kurchatov Institute", Gatchina,
188300, Russia}\affiliation{Clark Atlanta University, Atlanta, GA
30314, USA} \author{A. Z. Msezane}\affiliation{Clark Atlanta
University, Atlanta, GA 30314, USA}
\begin{abstract}
\end{abstract}
\maketitle

The authors of Ref. \cite{1} state: "Our hope, is that the topics
we have presented will provide inspiration for others working in
this field and motivation for the idea that significant progress
can be made on very hard problems if we focus our collective
energies. They continue: "We hope that this document can serve as a
starting point for further debate."

Taking their remarks into account, in our commentary we show that
some of the "very difficult problems" have been successfully
solved. We have to focus on some of the resolved problems, since
the authors claim: "Our hope, however, is that the topics we have
presented will provide inspiration for others working in this field
and motivation for the idea that significant progress can be made
on very hard problems if we focus our collective energies." Indeed,
there is no need to mislead potential researchers.

1. The authors claim: "The persistence of a linear resistivity at
the lowest temperatures when superconductivity is quenched is
perhaps one of the most difficult aspects to explain theoretically,
due to the lack of a scattering mechanism that gives $\tau\propto
1/T$ at temperatures lower than the Fermi energy, Debye
temperature, and other relevant energy scales." It is incorrect
statement, since this behavior was predicted many years ago and
confirmed later in our research \cite{2a}.

2. The authors state: "And ultimately, any microscopic theory
should give insight into why some metals are strange and some are
not." The answer to this passage is that the flat bands predicted
in 1990 \cite{2,3,4} are the fundamental building block of the
microscopic fermion condensation theory that can systematically
explain a huge number of experimental results, including the
non-Fermi liquid behavior, $\omega/T$ scaling, quantum spin
liquids, quasicrystals, etc. \cite{5,6}.

3. Regarding the correlated superconductors, the authors write: "To
complement and build upon these studies, we must seek new ways to
explore properties of both the normal and superconducting states to
begin addressing the questions outlined above." Experimental data
show that correlated superconductors and conventional ones have
common properties. They have the Bogoliubov quasiparticles and
exhibit common scaling behavior of the scaled condensation energy,
and satisfy the experimental Homes' law, see, for example,
\cite{homes,donald,prb2015,prlq,mat,npbq}. Overall, these scaling
relationships lead to the identification of fundamental laws of
nature and reveal the essence of superconductor physics, suggesting
that fermion condensation theory can be useful in that case, see
e.g. \cite{8}. In contrast to the traditional BCS theory of
superconductivity, in our case flat bands play significant role,
where the superconducting critical temperature $T_c$ becomes
proportional to the coupling constant $g$, $T_c\propto g$
\cite{2,4,5,7,8}; the single-particle spectrum of the corresponding
flat band strongly depends on the superconducting gap \cite{9}.
This was predicted in 2000 year \cite{5,10}, and confirmed by
experimental data \cite{12}.

In summary, clearly the authors \cite{1} have not reviewed
adequately the relevant literature. Given our comments above, their
encouraging statement "...it is clear that the Future of the
Correlated Electron Problem will be full of fascinating physics and
unexpected twists and turns that will challenge us for years to
come,"  would have been much more convincing and helpful if they
had given due consideration to the literature.


\begin{thebibliography}{99}

\bibitem{1} A. Alexandradinata, N. P. Armitage, A. Baydin, {\it et. al.},
SciPost Phys. Comm. Rep. {\bf 8} (2025),
https://doi.org/10.21468/SciPostPhysCommRep.8; arXiv: 2010.00584.

\bibitem{2a} M.V. Zverev, V.A. Khodel, and V.R. Shaginyan, JETP
Lett. {\bf 60}, 541 (1994); J. Dukelsky, V.A. Khodel, P. Schuck,
and V.R. Shaginyan,  Z. Phys. B{\bf 102}, 245 (1997); V. R.
Shaginyan, A. Z. Msezane, K. G. Popov, J. W. Clark, M. V. Zverev,
and V. A. Khodel, Phys. Rev. B {\bf 86}, 085147 (2012).

\bibitem{2} V.A. Khodel, V.R. Shaginyan, JETP Lett. {\bf 51}, 553 (1990).

\bibitem{3} G.E. Volovik, JETP Lett.  {\bf 53}, 222 (1991).

\bibitem{4} V.A. Khodel, V.R. Shaginyan, V.V. Khodel,
Phys. Rep. {\bf 249}, 1 (1994).

\bibitem{5} V.R. Shaginyan, M.Ya. Amusia, A.Z. Msezane, K.G. Popov,
Phys. Rep. {\bf 492}, 31 (2010).

\bibitem{6} M.Ya. Amusia, V.R. Shaginyan, Strongly Correlated Fermi
Systems: A New State of Matter. In Springer Tracts in Modern
Physics; Springer Nature: Cham, Switzerland, 2020; Volume 283.

\bibitem{homes} C.C. Homes, S.V. Dordevic, M. Strongin, D.A. Bonn,
R. Liang, W.N. Hardy, S. Komiya, Y. Ando, G. Yu, N. Kaneko, X.
Zhao, M. Greven, D.N. Basov, and T. Timusk, Nature {\bf 430}, 539
(2004).

\bibitem{prb2015} J.S. Kim, G.N. Tam, and G.R. Stewart,
Phys. Rev. B {\bf 92}, 224509 (2015).

\bibitem{prlq} A. Hunter, S. Beck, E. Cappelli, F. Margot,
M. Straub, Y. Alexanian, G. Gatti, M.D. Watson, T. K. Kim, C.
Cacho, N.C. Plumb, M. Shi, M. Radovic, D.A. Sokolov, A.P.
Mackenzie, M. Zingl, J. Mravlje, A. Georges, F. Baumberger, and A.
Tamai, Phys. Rev. Lett. {\bf 131}, 236502 (2023).

\bibitem{mat} H. Matsui, T. Sato, T. Takahashi, S.C. Wang, H.B. Yang,
H. Ding, T. Fujii, T. Watanabe, and A. Matsuda, Phys. Rev. Lett.
{\bf 90}, 217002 (2003).

\bibitem{npbq} K.J. Xu, Q. Guo, M. Hashimoto, Z.X. Li, S.D. Chen,
J. He, Y. He, C. Li, M.H. Berntsen, C.R. Rotundu, Y.S. Lee, T.P.
Devereaux, A. Rydh, D.H. Lu, D.H. Lee, O. Tjernberg, and Z.X. Shen,
Nat. Phys. {\bf 19}, 1834 (2023).

\bibitem{donald} A. Shekhter, M.K. Chan, R.D. MacDonald, and N.
Harrison, arXiv:2504.02179.

\bibitem{8} V.R. Shaginyan, M.Z. Msezane, S.A. Artomonov,
 JETP Lett. {\bf 122}, 158 (2025);

\bibitem{7} P. T\"orm\"a, S. Peotta, B.A. Bernevig,
Nat. Rev. Phys. 2022, {\bf 4}, 528.

\bibitem{9} V.R. Shaginyan, M.Z. Msezane, M.Ya. Amusia, G.S. Japaridze,
EPL 2022, {\bf 138}, 16004.

\bibitem{10} M.Ya. Amusia, V.R. Shaginyan, Phys.  Lett.  A {\bf 275}, 124 (2000).

\bibitem{12} W. Qin,  B. Zou, A.H. MacDonald, \prb {\bf 107}, 024509 (2023).


\end{thebibliography}
\end{document}